\documentclass[12pt, notitlepage]{article}
\pdfoutput=1
\usepackage{float}
\usepackage{subfig}
\usepackage{graphicx}
\usepackage[T1]{fontenc}
\usepackage[utf8]{inputenc}
\usepackage{authblk}
\usepackage{amsmath,bm}
\usepackage[left=3cm,right=3cm,top=2cm,bottom=2cm]{geometry}
\renewcommand\footnotemark{}

\usepackage{braket}

\usepackage{color}

\newcommand{\rf}[1]{(\ref{#1})}
\newcommand{\beq}{\begin{equation}}
\newcommand{\beql}[1]{\beq\label{#1}}
\newcommand{\eeq}{\end{equation}}
\newcommand{\bea}{\begin{eqnarray}}
\newcommand{\eea}{\end{eqnarray}}

\newcommand{\tr}{\mathrm{tr}\,}

\begin{document}

\title{{Signature Change of the Metric in CDT Quantum Gravity?}}

\author{J. Ambj\o rn$\,^{a,b}$}
\author{D.N.~Coumbe$\,^{c}$}
\author{J. Gizbert-Studnicki$\,^{c}$}     
\author{J. Jurkiewicz$\,^{c}$}
\affil{\footnotesize

$^a$~The Niels Bohr Institute, Copenhagen University\\
Blegdamsvej 17, DK-2100 Copenhagen \O , Denmark.\\
{Email: ambjorn@nbi.dk}\\

\vspace{10pt}

$^b$~IMAPP, Radboud University\\
 Niemegen, The Netherlands \\

\vspace{10pt}

$^c$~Faculty of Physics, Astronomy and Applied Computer Science, Jagiellonian University\\ 
ul. prof. Stanislawa Lojasiewicza 11, Krakow, PL 30-348, Poland.\\
{ Email: daniel.coumbe@uj.edu.pl, jakub.gizbert-studnicki@uj.edu.pl, jerzy.jurkiewicz@uj.edu.pl}\\
}


\date{\today}          
\maketitle

\begin{abstract}

We study the effective transfer matrix within the semiclassical and bifurcation phases of CDT quantum gravity. We find that for sufficiently large lattice volumes the kinetic term of the effective transfer matrix has
a different sign in each of the two phases. We argue that this sign change
can be viewed as a Wick rotation of the metric. We discuss the likely microscopic mechanism responsible for the bifurcation phase transition, and propose an order parameter that can potentially be used to determine the precise location and order of the transition. Using the effective transfer matrix we approximately locate the position of the bifurcation transition in some region of coupling constant space, allowing us to present an updated version of the CDT phase diagram.

\end{abstract}


\begin{section}{Introduction}

General relativity has already been successfully formulated as an effective quantum field theory that is valid up to some energy cut-off scale. Surprisingly, the work of Donoghue \cite{Donoghue:1995cz} and others even suggests that gravity forms the best perturbative quantum field theory in nature. However, as one increases the energy scale beyond the energy cut-off in a perturbative expansion new divergences appear that require an infinite number of counterterms  to define the theory, as first suggested in the seminal work of t' Hooft and Veltman \cite{'tHooft:1974bx} and later explicitly confirmed by Goroff and Sagnotti \cite{Goroff:1985th}. When considering small perturbations about flat Minkowski space one observes that the divergences cancel at the one-loop level. However, at the two-loop level and higher, such cancellations do not occur and divergences are once again present. The problem is compounded when one includes matter content, with nonrenormalizability occurring again at the one-loop level \cite{Donoghue:1997hx}. The so-called perturbative nonrenormalizability of gravity has led a number of researchers to investigate the idea that one should extend the idea of renormalization to the nonperturbative regime, which has become known as the asymptotic safety scenario, as first suggested by Weinberg \cite{Weinberg79}. 

If the asymptotic safety scenario is correct, gravity is effectively renormalizable when formulated nonperturbatively because the renormalization group flow of couplings end at a non-trivial fixed point in the high energy limit, and therefore remain finite over the entire range of energy scales. Evidence for such a fixed point has come mainly from functional renormalization group methods \cite{Reuter:2001ag,Lauscher:2001ya,Litim:2003vp,Codello:2007bd,Codello:2008vh,Benedetti:2009rx} and lattice approaches to quantum gravity \cite{Ambjorn05,Ambjorn:2007jv,Ambjorn:2008wc,Hamber:2009mt}. In a lattice formulation of quantum gravity a non-trivial fixed point would appear as a second-order critical point, the approach to which would define a continuum limit \cite{Ambjorn:1997di}. A lattice formulation of gravity is thus desirable: one can search for non-trivial fixed points by looking for a continuum limit of the theory, and at the same time one can perform calculations with controlled systematic errors. In addition it complements 
the analytic  renormalization group approach. 

A particular approach to lattice quantum gravity is defined by causal dynamical triangulations (CDT). In CDT, spacetime geometries are defined by locally flat $n$-dimensional simplices that are that glued together along their $(n-1)$-dimensional faces, forming a $n$-dimensional simplicial manifold. The defining characteristic of CDT is the introduction of a causality condition, in which one distinguishes between space-like and time-like links on the lattice. Hence, one defines a foliation of the lattice into spacelike hypersurfaces, each with the same fixed topology. Only geometries that can be foliated in this way are included in the ensemble of triangulations that define the path integral measure. The CDT approach to quantum gravity has enjoyed a number of successes, including the emergence of a 4-dimensional de Sitter-like geometry  \cite{Ambjorn:2007jv} and the likely existence of a second-order transition line in the coupling 
constant space. This second order transition line  may allow one to establish a continuum limit of the theory \cite{Ambjorn:2011cg}.     

Following the work of Regge \cite{Regge:1961px}, CDT discretises the continuous path integral and Einstein-Hilbert action into 

\begin{equation} \label{eq:CDTPartitionFunction}
Z_{E}={\sum_{T}}\frac{1}{C_{T}}e^{-S_{EH}}
\end{equation}

and

\begin{equation} \label{eq:GeneralEinstein-ReggeAction}
S^{Regge}_{EH}=-\left(\kappa_{0}+6\Delta\right)N_{0}+\kappa_{4}\left(N_{4,1}+N_{3,2}\right)+\Delta\left(2N_{4,1}+N_{3,2}\right),
\end{equation}

\noindent respectively. Here the CDT partition function of Eq. (\ref{eq:CDTPartitionFunction}) is defined as the sum over all possible triangulations $T$, and $C_{T}$ is a symmetry factor. The discretised Einstein-Regge action of Eq. (\ref{eq:GeneralEinstein-ReggeAction}) is defined in terms of the bare coupling constants $\kappa_{0}$, which is inversely proportional to Newton's constant, and $\Delta$ which is an asymmetry parameter defining the ratio of the length of space-like and time-like links on the lattice. There are two types of fundamental building blocks in CDT, the $\left(4,1\right)$ and $\left(3,2\right)$ simplices (see Ref. \cite{Ambjorn05} for a detailed discussion of the numerical setup), the number of which are quantified by $N_{4,1}$ and $N_{3,2}$, respectively. $N_{0}$ is the number of vertices in the triangulation $T$. $\kappa_{4}$ is formed from a linear combination of the cosmological and inverse Newtonian coupling constants, and is tuned to its (pseudo-)critical value such that one can take an infinite-volume limit. This leaves a parameter space that can be explored by independently varying the bare couplings $\kappa_{0}$ and $\Delta$.  

Typically one explores such a parameter space, trying to locate the position and order of its phase transitions, by studying a suitably defined order parameter depending on global properties of the triangulations, e.g. the number of vertices $N_{0}$, etc. However, changes of such an order parameter do not necessarily give much insight into the microscopic nature of the phase transition. Additional information regarding the microscopic properties of phase transition can be obtained by studying the effective transfer matrix linking the nearest (in integer time $t$) spatial slices \cite{Ambjorn:2012pp,Ambjorn:2014mra}.  
The transfer matrix $M$ and the associated effective Lagrangian  $L_{eff}$
\beq\label{TMLag}
\bra{n_{t+1}}  M \ket{n_{t}}\propto  \exp(-L_{eff}[n_t,n_{t+1}])
\eeq
 are parametrised by the spatial 3-volume observable $n_t\equiv N_{4,1}(t)$ which can be measured in Monte Carlo simulations. The existence of the
 effective action, parametrised by the pseudo-local form \rf{TMLag} is highly non-trivial. The effective Lagrangian contains a `kinetic' term coupling the
 neighbouring volumes and the diagonal `potential' term. Parameters of the action determine the phase structure of the model. In the original study three phases (denoted  A, B and C) were discovered, out of which the C phase, also called the de Sitter phase, was physically the most interesting, predicting the extended four-dimensional semi-classical background 
 geometry. Phase A was characterised by a lack of correlation between volumes in the neighbouring slices. In the B phase the time dependence of configurations was reduced to a single time slice.

It was argued in Ref. \cite{Ambjorn:2014mra} that the transition between phases A and C can be identified by the vanishing of the kinetic term of the effective transfer matrix (\ref{TMLag}). In  phase C the kinetic term is Gaussian with a positive coefficient.  Increasing the coupling $\kappa_{0}$ for fixed $\Delta$ one moves towards the A-C phase
transition and at the transition the coefficient multiplying the 
kinetic term vanishes. One can also perform the same study but keeping $\kappa_{0}$ fixed and varying $\Delta$. In this case one encounters a new so-called bifurcation phase separating phase C from phase B \cite{Ambjorn:2014mra}, see Fig. \ref{newphasediag}. Within this new bifurcation phase the kinetic term of the effective transfer matrix (\ref{TMLag}) bifurcates from a single Gaussian characteristic of small spatial volumes to a sum of two shifted Gaussians for large volumes. As we will argue in the following section  
 it is tempting to view this as evidence that the metric undergoes a signature change, such that one has a Lorentzian metric signature for sufficiently large $\Delta$ within phase C, but for sufficiently small $\Delta$ we encounter the bifurcation phase where the metric {effectively} changes to a Euclidean metric.

In a lattice theory of quantum gravity the hope is that one can take a continuum limit by approaching a second order critical point, at which one can take the lattice spacing $a\rightarrow 0$ whilst keeping observables fixed in physical units. In Ref. \cite{Ambjorn:2011cg} a likely second order transition was identified for $\Delta \sim 0$, thus raising the exciting possibility of defining a continuum limit in CDT. The newly reported bifurcation phase \cite{Ambjorn:2014mra}, however, exists in the parameter space between the physical phase C and the second order transition, and so being able to approach the second order transition from within the physical phase seems less likely. Therefore, it is important to  establish  the actual extent of the bifurcation phase in the CDT  parameter space and check whether phases C and B meet directly in some region. Alternatively, it is worth investigating whether the transition between phase C and the bifurcation phase is itself second order, thus raising the possibility of taking a continuum limit at an entirely new point in the parameter space.

\end{section}

\begin{section}{The Bifurcation Phase and the Signature Change}

We begin with a short reminder of previous results concerning the effective transfer matrix in phase C (also called the de Sitter phase) and inside the bifurcation phase.  The transfer matrix is defined as the transition amplitude from spatial volume $n$ at (discrete) time $t$ to the spatial volume $m$ at time $t+1$, integrating out all other degrees of freedom. 
{The existence of such an effective transfer matrix parametrised by a spatial volume observable is a highly non trivial conjecture. It is based on numerical results of CDT simulations with a varying length of the (periodic) proper time axis, $t_{tot}$.  Empirical probability distributions measured for different  $t_{tot}$ can be combined to calculate the transfer matrix elements $\braket{n|M|m}$ \cite{Ambjorn:2012pp}. This result does not depend on the choice of possible combinations of $t_{tot}$ used to determine $\braket{n|M|m}$, which has been explicitly checked {within all CDT phases}. This self-consistency check provides strong evidence that the above conjecture is true. 
It was  shown that the effective transfer matrix  inside the de Sitter phase  is perfectly consistent with other methods of measuring the effective action \cite{Ambjorn:2012pp}, and it can be used to replicate the average spatial volume profile and the shape of quantum fluctuations \cite{Ambjorn:2014mra} observed in this phase. The transfer matrix approach was also used to parametrise the effective action in other phases of CDT, and to analyse the phase transitions. This led to the discovery of a new bifurcation phase, where the transfer matrix can again be used to construct a simplified model that explains the observed narrowing of the spatial volume profile \cite{Ambjorn:2014mra}.}

It was shown in \cite{Ambjorn:2012pp} that inside phase C the {effective} transfer matrix can be accurately parametrised by 
\beql{TMC}
\braket{n|M_C|m} = \underbrace{\exp\Bigg[- \frac{1}{\Gamma} \frac{(n-m)^2}{ (n+m)}\Bigg] }_{\text{kinetic part}}  \underbrace{\exp\Bigg[ - \mu \left(\frac{n+m}{2}\right)^{1/3} + \lambda \left(\frac{n+m}{2}\right) \Bigg] }_{\text{potential part}} ,
\eeq
 which leads to a  discretised minisuperspace effective Lagrangian
 \beql{LagC}
 L_{C}[n,m] = \frac{1}{\Gamma} \frac{(n-m)^2}{n+m} + \mu \left(\frac{n+m}{2}\right)^{1/3} - \lambda \left(\frac{n+m}{2}\right) \ ,
\eeq
where $\Gamma$, $\mu$ and $\lambda$ are parameters related to the (effective) Newton's constant, the  size of the CDT universe and the cosmological constant, respectively. The dynamics of the spatial 3-volume  is therefore described by quantum fluctuations around the semi-classical de Sitter solution
\beql{deSitter}
\braket{n_t} \propto \cos^3(\alpha \cdot  t) .
\eeq

Inside the bifurcation phase the situation is quite different, and the measured transfer matrix takes the form \cite{Ambjorn:2014mra}
\begin{equation}
\braket{n|M_B|m}=
\label{TMB}
\end{equation}
$$
 =\left[ \exp \left( -\frac{1}{\Gamma}\frac{\Big((n-m) - c[n+m]\Big)^2}{n+m}\right) + \exp \left( -\frac{1}{\Gamma}\frac{\Big((n-m) + c[n+m]\Big)^2}{n+m}\right)\right] V[n+m] \ ,
$$
where $c[n+m] \to c_0 (n+m - s_b) $ for large volumes ($n+m \gg s_b$) and   $c[n+m] \to 0 $ for small volumes ($n+m \ll s_b$), and $V[n+m]$ is the potential part dependent on $n+m$. 
The value of $s_b$ (the so-called bifurcation point) provides a characteristic scale for which the system changes from phase C like behaviour (for small volumes) to a new type of behaviour (for large volumes), as the kinetic term in (\ref{TMB}) bifurcates from a single Gaussian to a sum of two shifted Gaussians. The strength of bifurcation depends on the parameter $c_0$. The values of the effective parameters $s_b$ and $c_0$ are functions of the  bare coupling constants $\kappa_0$ and $\Delta$, and the transition between the bifurcation phase and phase C is associated with the limits $s_b\to \infty$ and $c_0\to 0$, where Eq. (\ref{TMB}) transforms into Eq. (\ref{TMC}).

Starting within phase C and keeping $\kappa_0$ fixed while decreasing $\Delta$, for  some critical value $\Delta^c$ we observe a phase transition where  $c_0$ changes from zero to some positive value.  Our results show that the change is  smooth (the phase transition is most likely second or higher order).
{In this case one should  be able to define a continuous theory in the vicinity of the phase transition, even though it is not clear whether the CDT geometry deep inside the bifurcation phase is physically relevant. Therefore it is the  infinitesimal neighbourhood  of the phase transition which is of particular interest, and we will analyse it  in detail in the remainder of this section.}

Very close to the transition the $c_0$ parameter is  small  and $s_b$ is  large. At the same time $\Gamma$ is practically unchanged compared to phase C. 
Let us now consider the large volume limit, such that $n+m \gg s_b$ and $n-m \ll \Gamma /  2 c_o$.\footnote{This is the limit in which the spatial volume changes quite smoothly from slice to slice, i.e. $(n-m)/(n+m) \ll \Gamma / 2 c_0 s_b$, which in fact is the case when quantum fluctuations are relatively suppressed with increasing total volume.} In this case one can expand Eq. (\ref{TMB}) in powers of ${ 2 c_0}  (n-m)/ {\Gamma} $ to obtain
\begin{equation}
\braket{n|M_B|m}=
\label{TMBexp}
\end{equation}
$$
 = \exp \Bigg[ -\frac{1}{\Gamma}\left(1-\frac{2 c_0^2(n+m)}{\Gamma}\right)\frac{(n-m)^2} {(n+m)} - \frac{4}{3}\left(\frac{ c_0(n-m)}{\Gamma}\right)^4+...\Bigg]\exp\Bigg[-\frac{c_0^2}{\Gamma}(n+m)\Bigg]V[n+m] \ ,
$$
which, assuming that the potential term is only changed a small amount compared to phase C, leads to the following effective Lagrangian
 \beql{LagB}
 L[n,m] \approx \frac{1}{\Gamma} \left(1-\frac{2 c_0^2(n+m)}{\Gamma}\right) \frac{(n-m)^2}{n+m} + \mu \left(\frac{n+m}{2}\right)^{1/3} - \left(\lambda  -\frac{c_0^2}{\Gamma}\right)   \left(\frac{n+m}{2}\right),
\eeq
where we omit terms of power four and higher in our expansion parameter in Eq. (\ref{TMBexp}).
It is now clear that for spatial volumes  large enough ($n+m > \Gamma / 2 c_0^2 $) the kinetic term in Eq. (\ref{LagB}) effectively flips sign from positive to negative. This is exactly what one would expect if the metric undergoes a Wick rotation $t\to -i t$. Therefore, it is tempting to interpret the new phase transition as a sign of {an effective} signature change from Lorentzian metric in phase C to Euclidean metric in the bifurcation phase.\footnote{Note that in the CDT approach we are already working in a Wick rotated regime so here we interpret phase C with Euclidean metric in imaginary time as having Lorentzian signature in regular time, and vice-versa.  Alternatively, one can treat CDT as a realisation of the Hartle-Hawking Euclidean universe \cite{HartleHawkingWF}, with some specific topological restriction ($S^1\times S^3$), in this case the signature would change from Euclidean to Lorentzian when going from phase C to the bifurcation phase. This  interpretation may provide the mechanism of tunnelling from the Euclidean early universe to the Lorentzian Universe we live in.} In this context the {transition to the bifurcation phase may} gain some physical meaning. 

Interestingly, at least in this simple model, the effective signature is scale dependent.  The signature change occurs 
purely due to quantum fluctuations in spatial volumes. Quantum fluctuations in the small volume regime allow the volume to rise above the limit $n+m > \Gamma / 2 c_0^2 $, exposing the system to the bifurcation structure and causing the metric to change sign. If the system is in the large volume regime the opposite effect may occur. As a result it may be possible to observe fluctuations between different states, one corresponding to Lorentzian signature and the other to Euclidean signature. The possibility that the metric may change sign as a function of distance scale has also been proposed in Ref. \cite{Coumbe:2015zqa} based on much more general considerations. 

\end{section}

\begin{section}{The microscopic nature of the phase transition}
The concept of an effective transfer matrix, an effective Lagrangian of CDT and its relation to the signature change described above is based on the spatial volume observable, i.e. a very global property  of the underlying geometry. In this approach one disregards all  details of  geometric structures which form the spatial layers of  constant proper time. A more detailed analysis of  the geometry of  such layers is interesting for two reasons. Firstly, it's very likely that on the microscopic level there is a marked difference between the geometrical structure of the de Sitter and bifurcation phases. If this is the case, then one can try to quantify this difference and use it as an order parameter signalling the phase transition. Secondly, one can ask if a possible change in geometry (and a resulting signature change) is a global phenomenon, or whether it's only dependent on the local details of the triangulation. In this section we will answer both questions based on a preliminary  study of the geometry in the bifurcation phase compared to the geometry inside the de Sitter phase. More details are to follow in future publications.

Our first  observation concerns the behaviour of the  average curvature  of   individual spatial slices
\beql{Curv}
\bar R(t)\equiv\frac{\int d^3x\sqrt{g_{(3)}}R_{(3)} }{\int d^3x\sqrt{g_{(3)}}  },
\eeq
where $g_{(3)}$ and $R_{(3)}$ are the induced (spatial) metric determinant and the Ricci scalar in the spatial layer  at time $t$, respectively. This can be defined by a deficit angle
\beql{CurvD}
\bar R(t) =\frac{1}{N_3(t)} \sum_{l}\Big(2 \pi -  O(l)\cdot \theta\Big), 
\eeq
where the sum is taken over spatial links in time $t$, $O(l)$ denotes the order of a link (number of spatial tetrahedra sharing the link) and $\theta=\arccos(1/3)$ is a dihedral angle of an equilateral tetrahedron. As each spatial slice is built from such identical  tetrahedra (each one with 6 dihedral angles), one can express the sum in Eq. (\ref{CurvD}) by the total number of  spatial links $N_1(t)$ and the total number of spatial tetrahedra $N_3(t)$ at time $t$, leading to 
\beql{CurvDN}
\bar R(t)=2 \pi \frac{N_1(t)}{N_3(t)} - 6 \, \theta, 
\eeq
which can be numerically measured in the CDT simulations.
\begin{figure}[h!]
\centering
\scalebox{.6}{\includegraphics{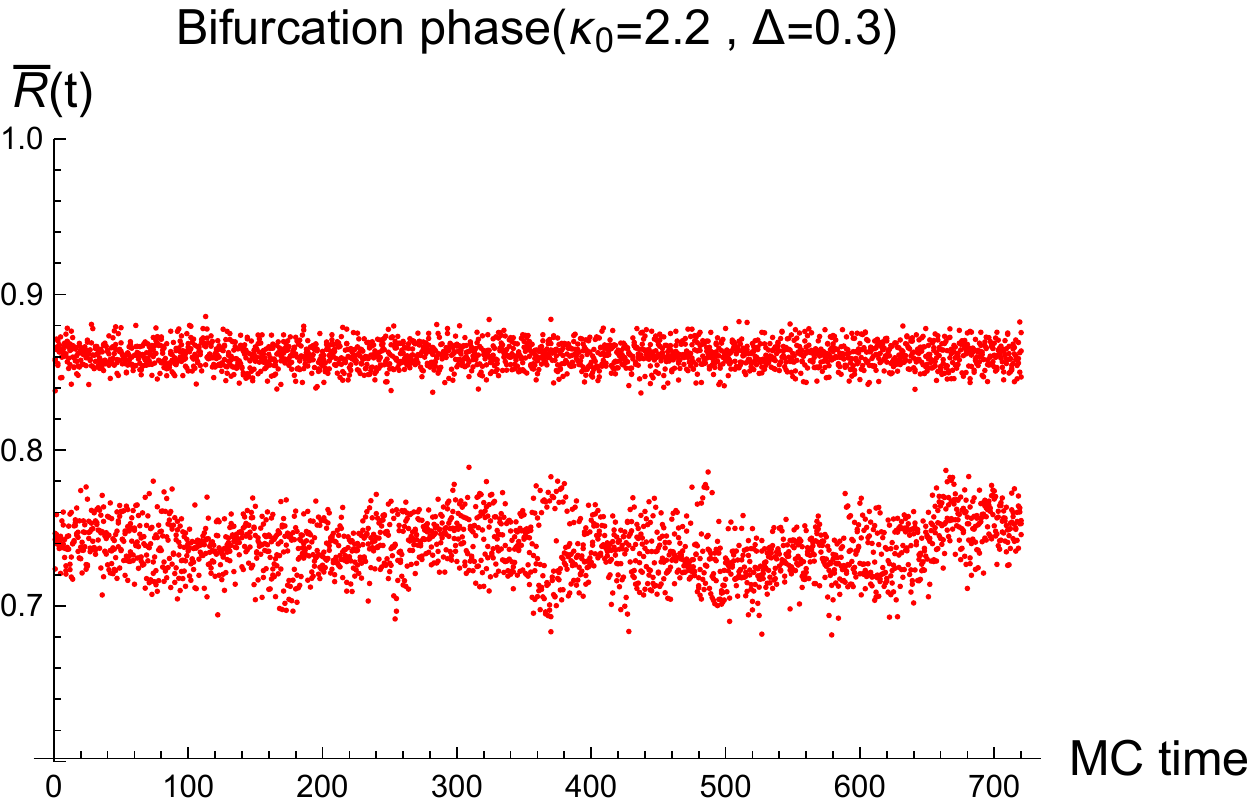}}
\scalebox{.6}{\includegraphics{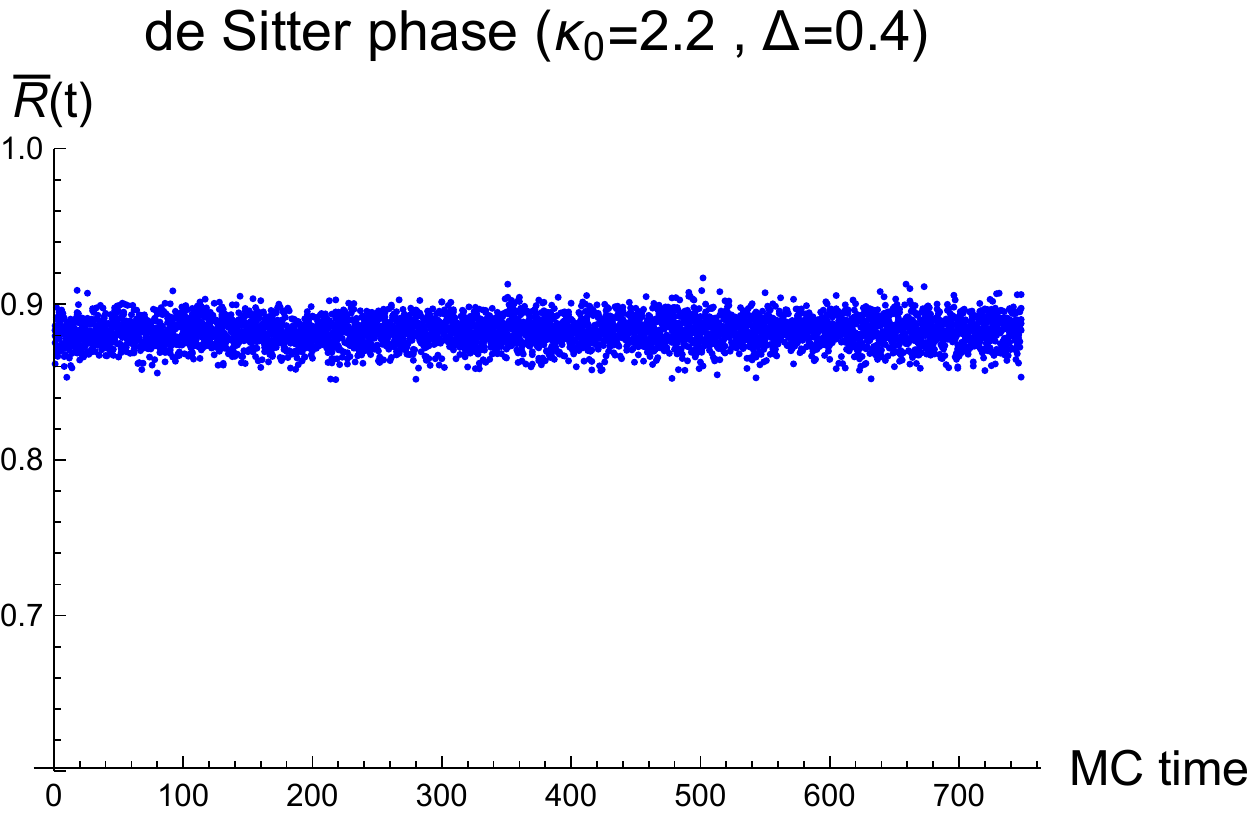}}
\caption{The average spatial curvature $\bar R(t)$ of Eq. (\ref{CurvDN})  inside the  bifurcation phase (left chart) and in  the de Sitter phase (right chart).  The horizontal axis represents Monte Carlo time, or alternatively single triangulations in which $\bar R(t)$ was measured for a range of $t$ from the central region of the blob (see Fig.\,\ref{FigAv} for details).
In the bifurcation phase $\bar R(t)$   is different for odd and even $t$, a difference that disappears in the de Sitter phase.}
\label{Rt}
\end{figure}
\begin{figure}[h!]
\centering
\scalebox{.7}{\includegraphics{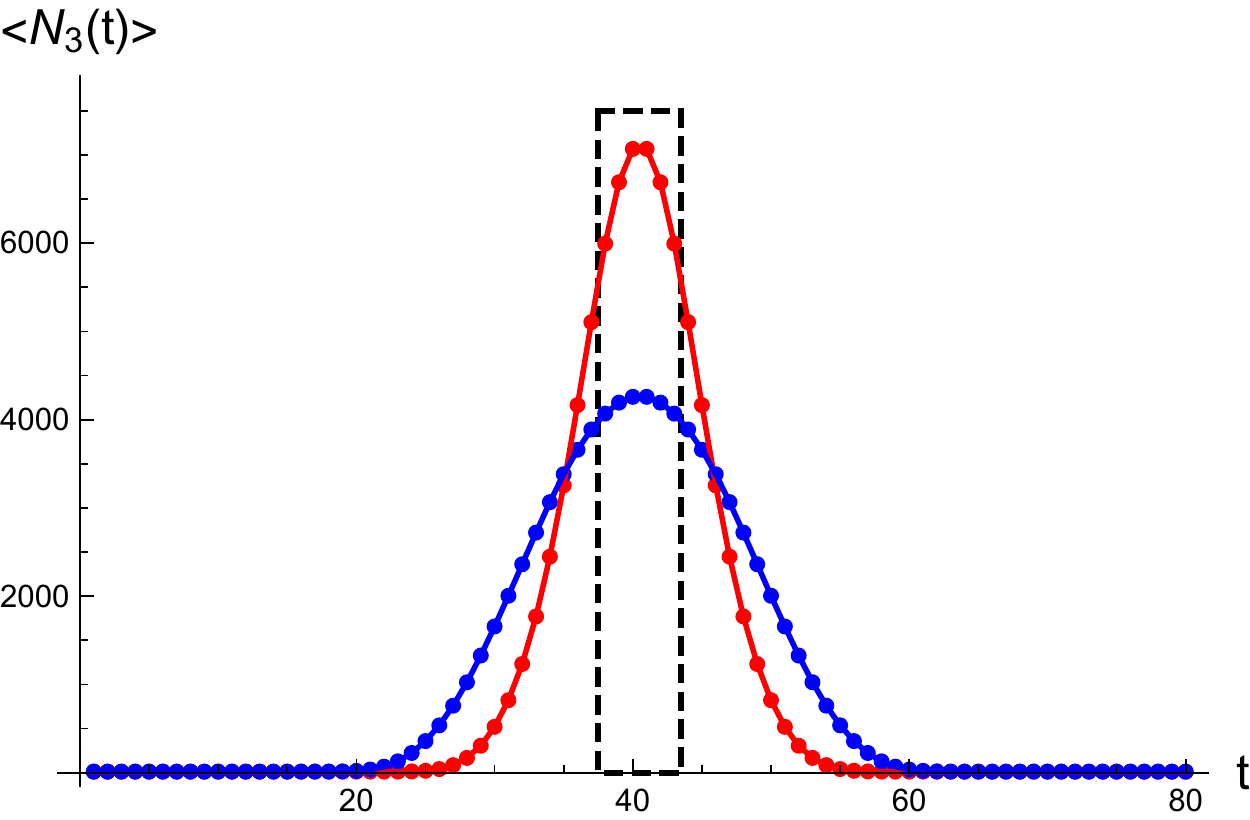}}
\caption{The spatial volume profile inside the bifurcation phase ($\Delta=0.3$) - red, and inside the de Sitter phase ($\Delta=0.4$) - blue,  for $\kappa_0=2.2$. The average $\braket{.}$ is taken over an ensemble of triangulations. In both cases $\braket{N_3(t)}$ is a smooth function of $t$ and both profiles look qualitatively the same. The dashed rectangle highlights  the central region  of  the blob where the average spatial curvature  $\bar R(t)$ was measured for odd and even $t$ (see Fig.\,\ref{Rt}).}
\label{FigAv}
\end{figure}

The average spatial curvature $\bar R(t)$ measured for a choice of the bare coupling constants ($\kappa_0=2.2$, $\Delta=0.3$) inside the  bifurcation phase   jumps between two different values observed for odd and even $t$, respectively (see Fig.\,\ref{Rt} left). The jumps in  $\bar R(t)$ happen despite the fact that the spatial volume profile $\braket{N_3(t)}$ is itself a smooth function of $t$, and at least qualitatively does not differ significantly from the profile inside the de Sitter phase (see Fig.\,\ref{FigAv}).
This `anti-ferromagnetic' like behaviour of $\bar R(t)$ smoothly vanishes as one increases $\Delta$ while keeping $\kappa_0$ fixed, and completely vanishes inside the  de Sitter phase (above $\Delta = 0.4$), where $\bar R(t)$ is constant from slice to slice  (see Fig.\,\ref{Rt} right).   Therefore, the phase transition can be signalled  by the following order parameter
\beql{OP1}
OP_1 = \left| \bar R(t_0) - \bar R( t_0+1) \right|, 
\eeq
where we define the (integer) time $t_0$ to be closest to  the centre of volume of a triangulation.\footnote{In our approach the discrete centre of volume $t_0$ is defined up to one time slice, therefore to calculate $OP_1$ we average over 3 values of  $\left| \bar R(t) - \bar R( t+1) \right|$ calculated for $t=t_0-1$, $t=t_0$ and $t=t_0+1$.} The plot of  $\braket{OP_1}$ as a function of $\Delta$ for fixed $\kappa_0=2.2$ can be found in Fig.\,\ref{OP} (left), where one can observe  the phase transition for $\Delta^c \approx 0.35$. 
\begin{figure}[h!]
\centering
\scalebox{.6}{\includegraphics{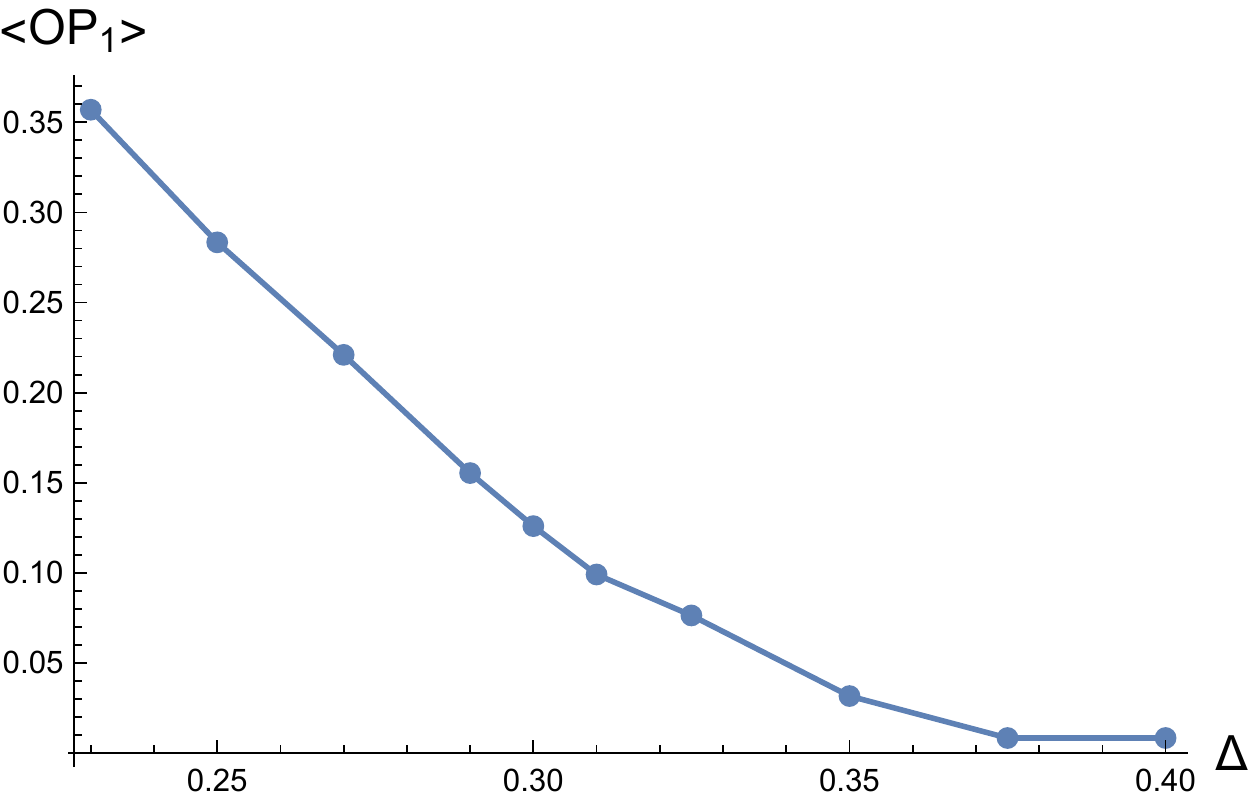}}
\scalebox{.6}{\includegraphics{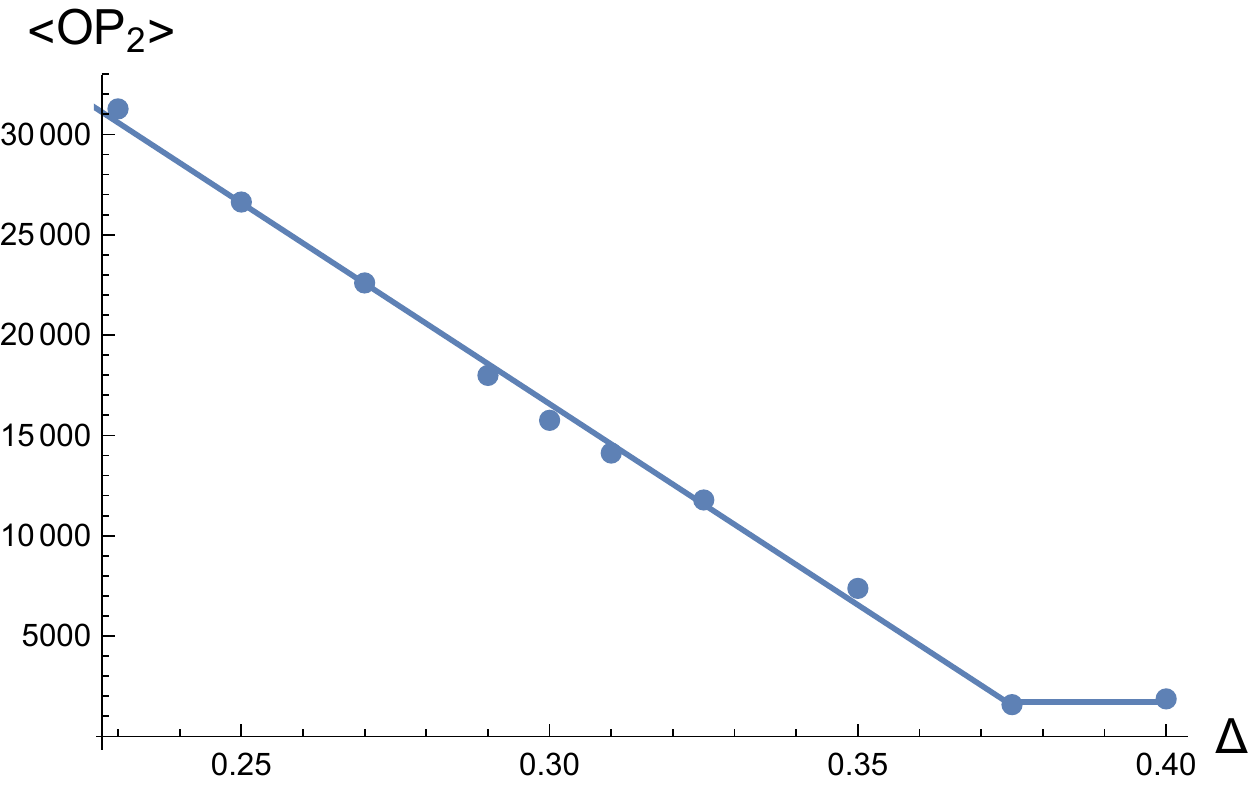}}
\caption{Order parameters $OP_1$ (defined in Eq. (\ref{OP1})) - left chart, and $OP_2$ (defined in Eq. (\ref{OP2})) - right chart, measured for a range of $\Delta$ values for fixed $\kappa_0=2.2$. The average $\braket{.}$ is taken over the ensemble of triangulations. Both order parameters point to the phase transition around $\Delta^c\approx 0.35.$}
\label{OP}
\end{figure}

The fluctuations of $\bar R(t)$ between odd and even time slices in the bifurcation phase should be related to differences in  their geometry  and one can try to analyse this in more detail. The spatial geometry of each slice is encoded in the connectivity of its building blocks and depends on the number of sub-simplices (triangles, links and vertices) shared by the neighbouring tetrahedra. As an example, consider the most basic building block, the vertex. If we take into account a  topological constraint related to spherical topology of spatial slices $N_0(t)-N_1(t)+N_3(t)=0$, where $N_0(t)$ is the number of vertices in time $t$, Eq. (\ref{CurvDN}) can be written as
\beql{CurvDNo}
\bar R(t)= 2\pi \frac{N_0(t)}{N_3(t)} - \text{const} \quad,\quad \text{const}=6 \, \theta - 2\pi > 0 .
\eeq
Therefore, the  observed jumps in $\bar R(t)$ are related to differences in the  number of vertices shared by  tetrahedra, forming odd and even spatial slices of generic triangulations. 

Let us look in detail at  two  neighbouring  slices with high and low average spatial curvature $\bar R(t)$, respectively. Eq. (\ref{CurvDNo}) suggests that the average coordination number of a vertex $O(v)$ (the number of 4-simplices which share the vertex) should differ depending on whether we look at odd or even slices. Our preliminary results show that the difference is mainly caused by just  one 'singular' vertex \footnote{The coordination number of such a vertex is typically a few orders of magnitude higher than the average coordination number in the slice.} present in each slice with high $\bar R(t)$, and  not present in slices with low $\bar R(t)$. Inside the de Sitter phase the situation is vastly different as such 'singular' vertices are not present at all.
One can therefore define another order parameter  based on the difference in maximal  coordination number of vertices in odd and even spatial slices\footnote{Here we define $t_0$ as the slice with a vertex with maximal coordination number in the whole 4-dimensional triangulation. This agrees very well with a centre of volume definition used in Eq. (\ref{OP1}). We also average over three values of $OP_2$ as described in footnote\,$^3$.}
\beql{OP2}
OP_2 = \Big|  \text{max}\big[O\big(v(t_0\big) \big] - \text{max}\big[O\big(v(t_0+1\big) \big] \Big| .
\eeq
This order parameter seems to change linearly for $\Delta < \Delta^c $ and it again suggests $\Delta^c \approx 0.35$ (for $\kappa_0=2.2$) - see Fig.\,\ref{OP} (right).  

Let us, at least qualitatively, translate  these results into a geometric language. The appearance of singular vertices in the bifurcation phase indirectly suggests that  a large fraction of  total volume  is concentrated within a short geodesic distance, forming `clusters' within a generic triangulation. This is  confirmed by direct analysis of the geometry around such singular vertices. Due to the presence of these clusters such geometries lack the homogeneous features of the de Sitter phase.\footnote{Due to discretisation and the fractal nature of  triangulations  one can only expect homogeneity on sufficiently large scales. It was shown in \cite{Cooperman:2014rha}  that   this is actually the case in the de Sitter like phase of 2+1 dimensional CDT. The same is being  verified in 3+1 dimensions and the  results will be published soon.} As a result the phase transition from the de Sitter  to the bifurcation phase is related to a spontaneous breaking of the translational symmetry of triangulations in spatial directions.\footnote{In the de Sitter phase configurations may be viewed as quantum fluctuations of a regular semi-classical background geometry, for which  the translational symmetry of the bare action (\ref{eq:GeneralEinstein-ReggeAction}) in the time direction is explicitly broken. This is also the case  in the bifurcation phase and additionally, in the same sense, the action is also broken in the spatial direction.}  We observe  that the coordination number of singular vertices grows when one goes deeper and deeper inside the bifurcation phase. In other words, the clusters grow in size `eating up' the rest of the triangulation. Eventually, this leads to another phase transition to the generic phase B, observed for low values of $\Delta$, where the whole triangulation consists of just one huge cluster (see phase diagram in Fig.\,\ref{newphasediag}).

It is also worth mentioning that in the bifurcation phase clusters of spatial volume that are closest to each other in  time (say the ones located in $t-1$ and $t+1$) are also linked to each other in  space, in a sense  that they share the same singular vertex in $t$. Such structure is repeated  periodically every second time slice (this is schematically illustrated in Fig.\,\ref{Grains}). As a result all clusters are connected in the time direction to form a kind of  `tube' embedded in the `sea' of  a different (probably similar to the de Sitter phase) geometry. We 
{suppose} that all  effects related to the possible signature change  discussed in the last section are due to the geometry of the `tube' and not to the `sea', and in this sense the signature change 
{might be a local phenomenon, triggered by large fluctuations of a (local) conformal factor. Similar effects were observed in two-dimensional 
quantum gravity interacting with conformal matter above the $c=1$ barrier}. 
However, determining the validity of this conjecture requires further study. 
\begin{figure}[h!]
\centering
\scalebox{.8}{\includegraphics{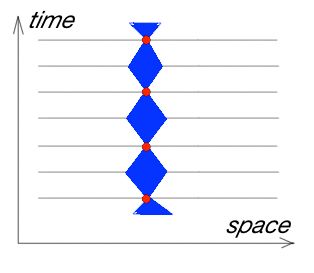}}
\scalebox{.59}{\includegraphics{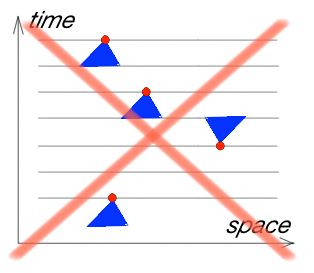}}
\caption{Schematic  of the geometric structure of the bifurcation phase.  Spatial volume is concentrated in clusters (blue). All clusters are connected by vertices (red) of  extremely high coordination number.  Such structures are repeated periodically, resembling an irregular `tube' embedded in the `sea' of different geometry.}
\label{Grains}
\end{figure}

\end{section}
\begin{section}{New phase diagram}

In  \cite{Ambjorn:2014mra} we   identified the   new bifurcation  phase and determined an approximate position of just one phase transition point in the phase diagram (see Fig. \ref{newphasediag}). These results were based on the study of the effective transfer matrix measured  for one  fixed value of the  bare coupling constant $\kappa_0=2.2$ and a range of $\Delta$. For this  choice of $\kappa_0$ the phase transition takes place  within the range $\Delta = 0.3 - 0.4$, which is consistent with the behaviour of the order parameters defined in last section (see Fig.\,\ref{OP}). 

We would now like to present the results of a more systematic study of the phase diagram. It is based  on measurements of the effective transfer matrix for a grid of points in the ($\kappa_0, \Delta$) bare coupling plane.\footnote{Here we used the transfer matrix method as a first estimate of the position of the phase transition line as it is computationally very efficient.  Work is in progress to use the order parameters defined in the previous section to get a very precise position of the phase transition.  Such an approach is more accurate but also much more computationally costly. The results will be presented in future publications.} Technically, this was done by performing  Monte Carlo simulations of CDT geometries with a very short length of the (periodic) proper time axis - just two spatial slices. In this case one can measure a probability distribution $P(n_{1},n_{2})$ of finding a spatial volume $n_1$ at time $t=1$, and $n_2$ at time $t=2$. In the effective transfer matrix approach this probability is given by
\beql{P2T2}
P(n_{1},n_{2}) = \frac{ \ {\braket{n_1 | M | n_{2}} \braket{n_{2} | M | n_1} }}{{\tr M^2}}\ ,
\eeq
and one  can use it to compute the transfer matrix elements.  Up to a normalisation factor one obtains
\beql{MT2}
\braket{n|M|m}= \sqrt{P(n_{1}=n,n_{2}=m)} \ .
\eeq
More  technical details can be found in \cite{Ambjorn:2012pp,Ambjorn:2014mra,JstudThesis}.

To study  the bifurcation transition we will focus on  selected empirical  transfer matrix cross-diagonals, i.e. elements for fixed $n+m =  s$. From Eqs. (\ref{TMC}) and (\ref{TMB}) one obtains
\beql{TMCross}
\braket{n|M|s-n}=
\eeq
$$
 =V[s] \left[ \exp \left( -\frac{\Big((m-n) - c[s]\Big)^2}{\Gamma s}\right) + \exp \left( -\frac{\Big((m-n) + c[s]\Big)^2}{\Gamma s}\right)\right] \ ,
$$
where the bifurcation shift $c[s]$ is positive within the bifurcation phase and is null in the de Sitter phase, and the potential part $V[s]$ turns  into a  normalisation factor. All parameters in Eq. (\ref{TMCross}) are of course functions of the bare couplings $\kappa_0$ and $\Delta$, and the phase transition is signalled by $c[s]\to 0$ .

In order to check how the phase transition depends on the size of the system we measured a number of cross-diagonals for $s=n+m = 10$k, $20$k, $30$k, $40$k, $60$k. All measurements were performed within the parameter ranges $\kappa_0=1.0 - 4.6$ and $\Delta = 0.0 - 0.4$ (in total we measured over $800$ cross-diagonals). The results presented below are still mostly approximate. To adopt a more accurate approach, i.e. to estimate a precise position of the new phase transition line and its dependence on the total volume, one would need to perform very dense  measurements close to the  transition. This is not an easy task as our Monte Carlo algorithm  looses efficiency for runs in this region of parameter space. This feature is   characteristic of numerical simulations close to transitions of second or higher order, where very long autocorrelation times occur. Therefore,  to assure the data is fully thermalized one needs to  increase the simulation time considerably. 

\begin{figure}[h!]
\centering
\scalebox{1.1}{\includegraphics{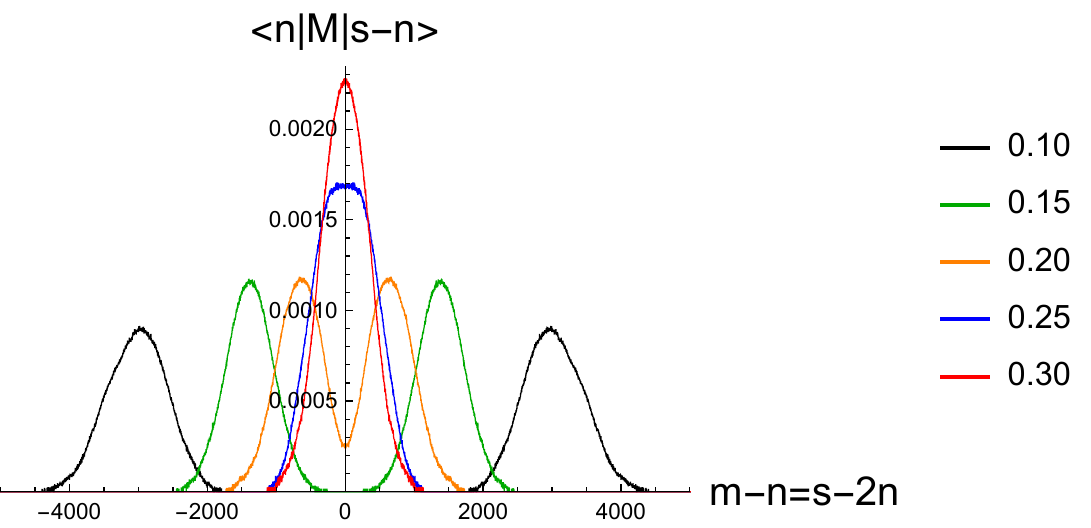}}
\caption{Empirical cross-diagonals measured for fixed $\kappa_0=2.2$ and a choice of $\Delta$ (denoted by different colours). Data is measured for $s=30$k.  A gradual vanishing of the bifurcation structure is visible when $\Delta$ is increased.  One can identify that the phase transition occurs within the range $\Delta=0.25-0.3$.}
\label{BifantiK}
\end{figure}
\begin{figure}[h!]
\centering
\scalebox{1.1}{\includegraphics{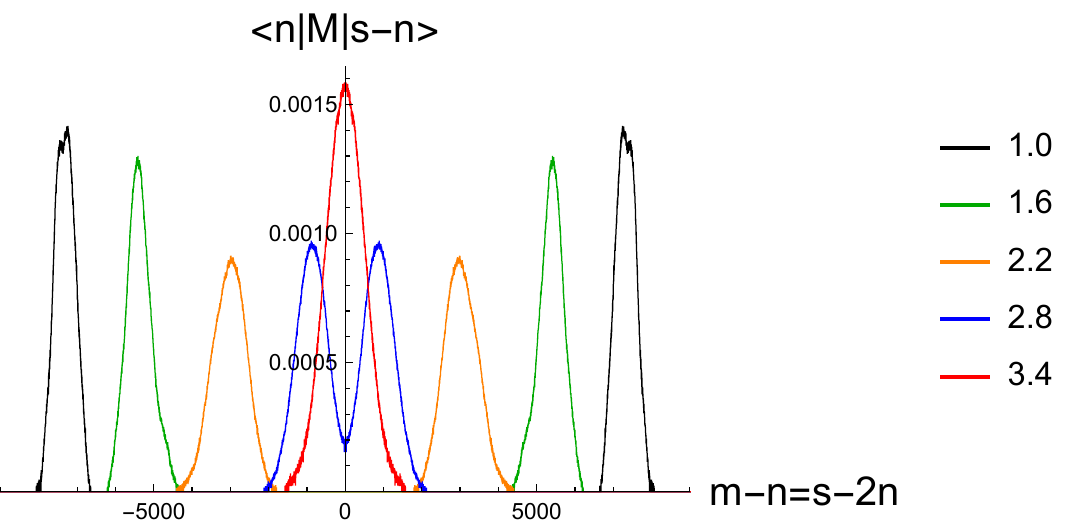}}
\caption{Empirical cross-diagonals measured for fixed $\Delta=0.1$ and a choice of $\kappa_0$ (denoted by different colours). Data measured for $s=30$k. A gradual vanishing of the bifurcation structure is visible when $\kappa_0$ is increased.  One can identify the phase transition to take place within the $\kappa_0=2.8-3.4$ range.}
\label{BifantiD}
\end{figure}
\begin{figure}[h!]
\centering
\scalebox{.6}{\includegraphics{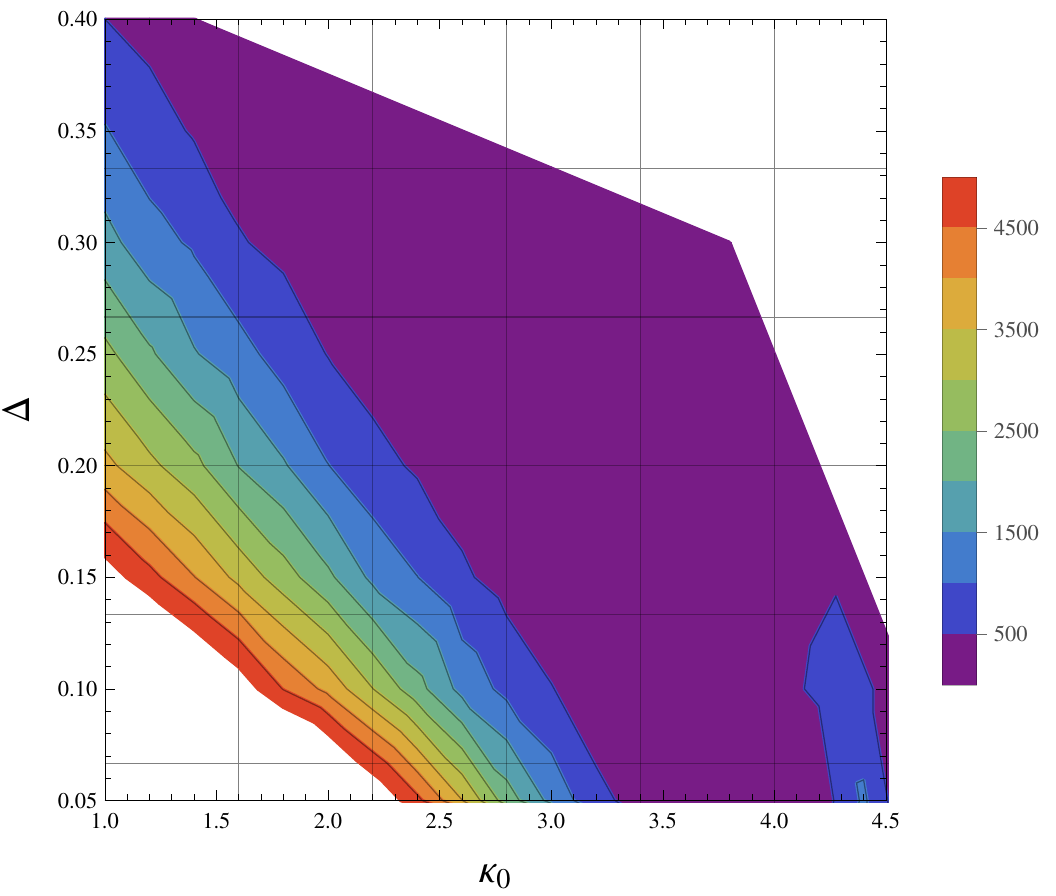}}
\scalebox{.6}{\includegraphics{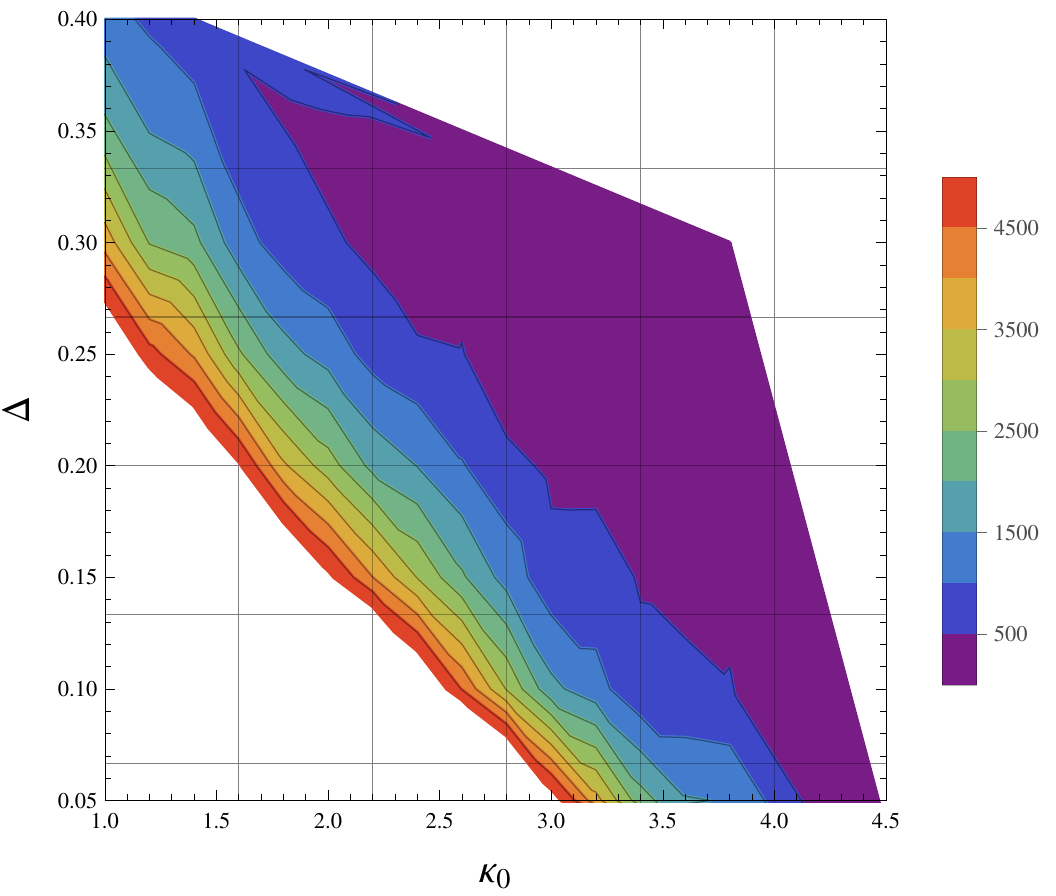}}
\caption{Contour plots of the bifurcation shift $c[s]$ in the ($\kappa_0,\Delta$) plane measured for $s=30$k (left) and $s=60$k (right). The phase transition seems to be shifted to the top-right when the total volume is increased.}
\label{Bifsiatka}
\end{figure}
\begin{figure}[h!]
\centering
\scalebox{.8}{\includegraphics{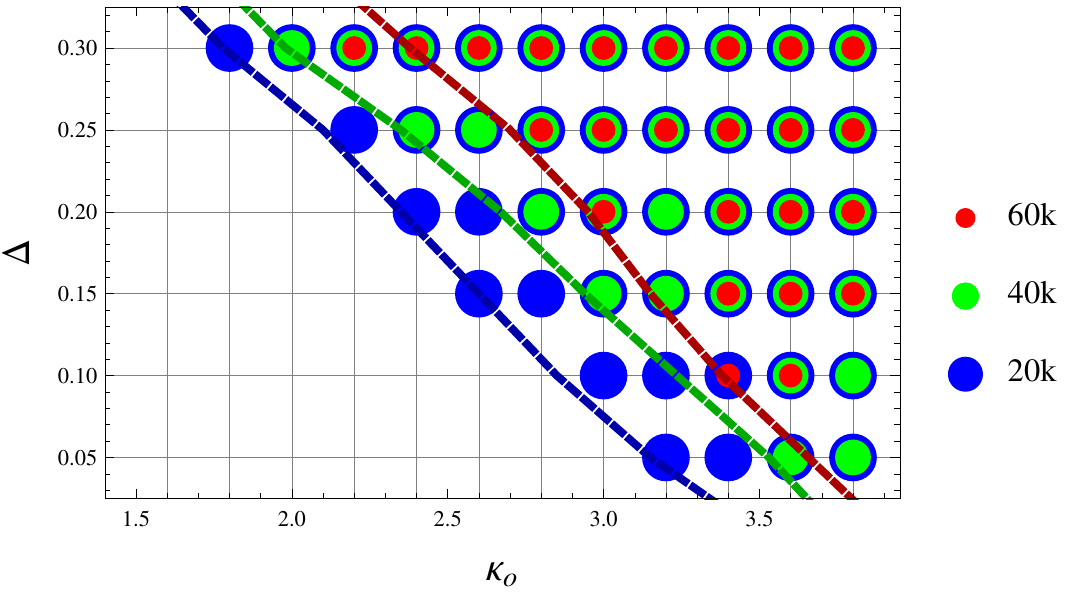}}
\caption{Points in the ($\kappa_0,\Delta$) bare coupling plane for which the bifurcation structure disappears (bifurcation shift $c[s]<100$),  measured for  $s= 20$k (blue), $40$k (green) and $60$k (red). The bottom-left edge of the dotted  regions can be associated with the new phase transition (dashed lines) measured for different  total volumes $s$. The red point visible for ($\kappa_0=2.2,\Delta=0.3$) was manually excluded due to a  small but still visible bifurcation structure.}
\label{BifsiatkaC}
\end{figure}
The measured cross-diagonals for fixed
$\kappa_0=2.2$ and a range  of $\Delta$  are presented in Fig. \ref{BifantiK}, and for fixed $\Delta=0.1$ and different $\kappa_0$'s in  Fig. \ref{BifantiD}. 
One observes a gradual vanishing of the bifurcation with increasing $\Delta$ and with increasing $\kappa_0$, respectively. This tendency is illustrated in Fig. \ref{Bifsiatka} where we present a contour plot of the measured bifurcation shift $c[s]$  in the ($\kappa_0, \Delta$) bare coupling plane. The left chart presents the data measured for the total volume  $s=30$k, and the right chart for $s= 60$k. The purple colour indicates a region of  vanishing bifurcation ($c[s] < 500$), which can be associated with phase C, while different colours denote higher values of $c[s]$ inside the bifurcation phase. The closer one approaches the  phase transition   the harder it is to see a non-zero bifurcation shift $c[s]$. This is caused by the  rising value of the bifurcation point $s_b\to \infty$ at the phase transition. As one can observe a bifurcation only for large spatial volumes ($s=n+m > s_b$), one can measure $c[s]>0$ only for very large $s$ close to the transition.  As a result the phase transition line seems to shift up and to the right in the ($\kappa_0, \Delta$) plane  as one increases $s$ - see Fig. \ref{BifsiatkaC}, where the coloured  dots denote the points inside the de Sitter phase and different colours   correspond to  different total volumes $s = 20$k (blue), $40$k (green) and $60$k (red). The bottom-left edge of the dotted regions can be associated with the phase transition line measured for different  values of $s$.  These results suggest that the de Sitter phase shrinks in favour of the bifurcation phase  as one increases the total volume and potentially its existence could be just a finite size effect. This scenario  cannot be completely excluded,  however the detailed studies of the geometry of both regions of the parameters space presented above as well as other observables, e.g. the behaviour of the spectral dimension for different total volumes, show that it is unlikely. All results suggest that phase C persists in the infinite volume limit and the real phase transition  is just very close to what we measure in the transfer matrix data for $s=60$k (the biggest total volume in our measurements).\footnote{All these results will be presented in a separate article that will follow this work.} We use this data to update the CDT phase diagram with a new  phase transition line - see Fig. \ref{newphasediag}. The line has been extrapolated both to the top-left and to the bottom-right, where we conjecture that all four phases meet at a common point,  becoming a quadruple point.  
\begin{figure}[h!]
\centering
\scalebox{.85}{\includegraphics{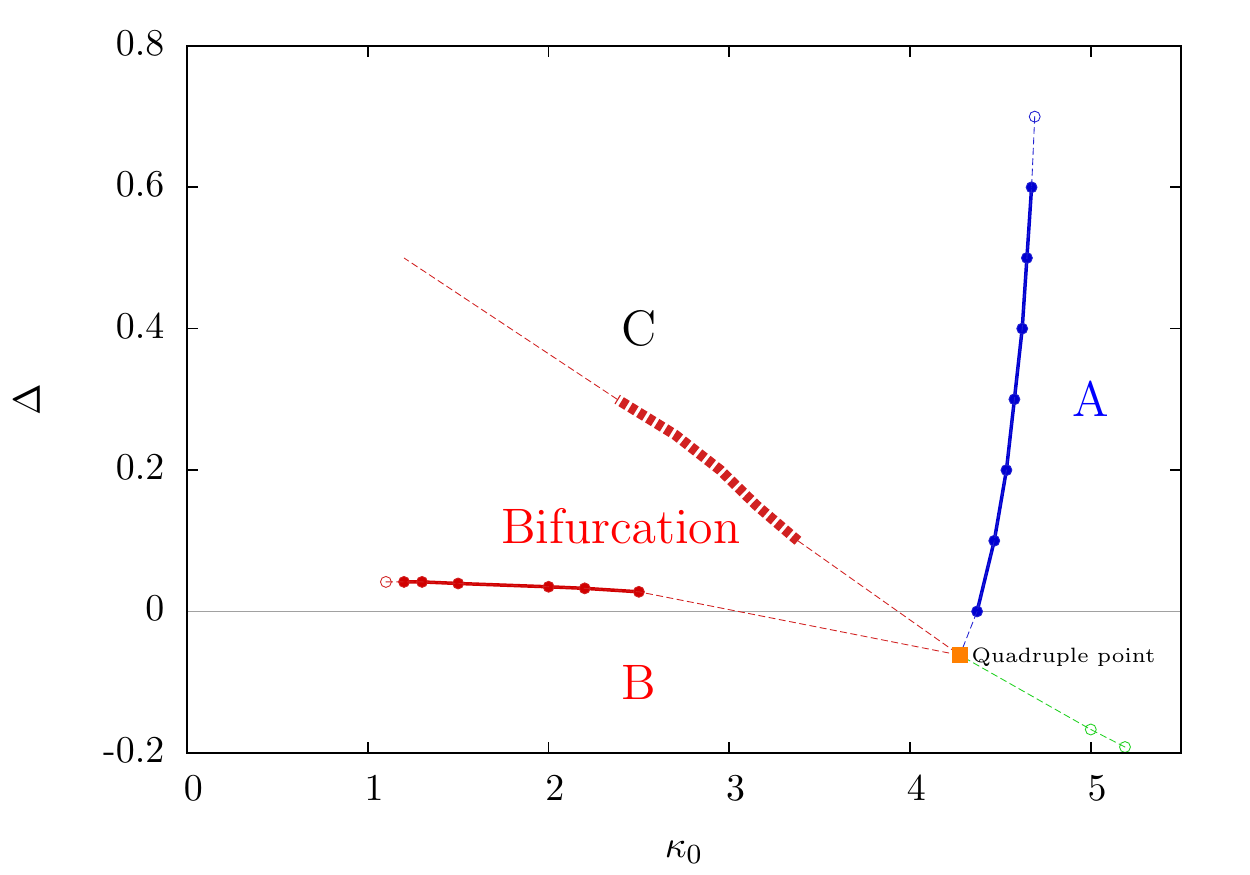}}
\caption{The updated  phase diagram of four-dimensional CDT. The approximate position of the phase transition between phase C and the new bifurcation phase (measured from the transfer matrix data for $s=60$k) is denoted by the thick dashed red line. The thin dashed line is an extrapolation. }
\label{newphasediag}
\end{figure}

\end{section}

\begin{section}{Discussion and Conclusions}

It is well known that the Wick rotation is an extremely useful mathematical trick, but it was not until the seminal work of Hartle and Hawking \cite{HartleHawkingWF} that anybody thought it could have physical relevance. Hartle and Hawking proposed that the spacetime metric might undergo a discontinuous Wick rotation in the early universe, thereby smoothing out the problematic big bang singularity and defining a very simple boundary condition for the universe, namely that there is no boundary \cite{HartleHawkingWF}. However, an underlying explanation for why the early universe might behave in such a way was notably absent. Since then, there have been a small number of similar proposals and possible explanations, most notably from loop quantum gravity 
and loop quantum cosmology {\cite{Cailleteau:2011kr,Mielczarek:2012pf,Barrau:2014maa}}. Independent of any particular approach to quantum gravity, Ref. \cite{Coumbe:2015zqa} also finds that spacetime appears to undergo a Wick rotation, but with the distinct difference that time continuously Wick rotates as a function of scale, and with the additional feature of scale dependent time dilation. This work presents the first evidence within the context of CDT quantum gravity that the metric appears to undergo a scale dependent Wick rotation, thereby providing numerical evidence in support of these more analytical results. 

We have studied the behaviour of the effective transfer matrix within the newly discovered bifurcation phase and within the established de Sitter phase of CDT. We find that for sufficiently large spatial volumes the kinetic term of the effective transfer matrix flips sign from positive to negative when crossing the transition between the de Sitter phase and the bifurcation phase. The natural interpretation of this is that the metric undergoes a Wick rotation $t\rightarrow -it$, transforming from Lorentzian signature in the de Sitter phase to Euclidean signature in the bifurcation phase {(see footnote 2)}. In this scenario, the presence of the newly discovered bifurcation phase may {have} a physical interpretation: the boundary of the bifurcation phase defines the points at which the metric changes between having Lorentzian and Euclidean signature. This highlights the importance of determining the order of the transition dividing the de Sitter and bifurcation phases, since a first order transition would suggest a discontinuous Wick rotation, whereas a higher order transition would allow for a smooth continuous Wick rotation. Hence, this result may be able to definitively rule out some of the models of signature change discussed above.   

A picture of the likely microscopic mechanism underlying the nature of the bifurcation transition is presented. As we probe deeper and deeper into the bifurcation phase we observe the formation of vertices with increasingly high coordination number, which are absent in the de Sitter phase. The formation of  dense clusters of simplices around these vertices results in the breaking of translational symmetry invariance in the spatial direction (see footnote 7), leading to a geometry that does not share the homogeneous properties of the de Sitter phase. It is the accumulation of these clusters that seems to be responsible for the distinctly different geometric properties of the de Sitter and bifurcation phases. {It is important to realise that the phase
structure and physical properties of systems analysed in the CDT model result from the balance between the physical Hilbert-Einstein action and the entropy  of geometric configurations. Within phase C, although the effective action has a form of the mini-superspace action  of Hartle and Hawking \cite{HartleHawkingWF} it has the opposite sign than that obtained in the original derivation, where all geometric degrees of freedom, except for the (Euclidean) time-dependent scale factor were excluded. This means that the instability of the conformal factor gets stabilised by entropy. The competition between the two effects is responsible for phase transitions, in particular for creating a bifurcation phase, where we observe large local fluctuations of   volume, which may be interpreted as a local dominance of the physical action over the entropy. This effect may be viewed analogously to the 
formation of large spin clusters near the phase transition between the disordered and ordered phases in the Ising model.}  
Based on the appearance of these clusters we propose an order parameter that can potentially be used to determine the precise location and order of the bifurcation phase transition. 

We use the effective transfer matrix to locate the approximate location of the bifurcation phase transition for multiple values of the bare coupling constants. This study allows us to present a new and updated picture of the CDT phase diagram of 4-dimensional CDT.

\end{section}

\section*{Acknowledgements}

D.N.C and J.J acknowledge the support of the grant DEC-2012/06/A/ST2/00389 from the National Science Centre Poland (NCN). J.G-S acknowledges the  National Science Centre Poland (NCN) support via the grant DEC-2012/05/N/ST2/02698. J.A was supported by the ERC-Advanced grant 291092,
``Exploring the Quantum Universe'' (EQU).



\bibliographystyle{unsrt}
\bibliography{Master}



\end{document}